\documentclass[conference]{src-asilomar/IEEEtran}
\IEEEoverridecommandlockouts

\usepackage{amsmath,amssymb,amsfonts}
\usepackage{algorithm}
\usepackage{algpseudocode}
\usepackage{graphicx}
\usepackage{siunitx}
\usepackage{textcomp}
\usepackage[nolist]{acronym}
\usepackage[super]{nth}
\usepackage{multirow}
\usepackage{arydshln}
\usepackage{hyperref}
\hypersetup{
  colorlinks=false,
  pdfborder={0 0 0}
}
\usepackage{cleveref}

\usepackage{multirow}
\usepackage{stfloats}
\usepackage{booktabs}
\usepackage[style=ieee]{biblatex}
\usepackage{graphicx}
\usepackage{wrapfig}
\ifCLASSOPTIONcompsoc
\usepackage[caption=false,font=normalsize,labelfont=sf,textfont=sf]{subfig}
\else
\usepackage[caption=false,font=footnotesize]{subfig}
\fi
\usepackage{tikz}

\usepackage{xcolor}

\def\mathlette#1#2{{\mathchoice{\mbox{#1$\displaystyle #2$}}%
		{\mbox{#1$\textstyle #2$}}%
		{\mbox{#1$\scriptstyle #2$}}%
		{\mbox{#1$\scriptscriptstyle #2$}}}}
\def\ve#1{\mathlette{\boldmath}{#1}}

\def\BibTeX{{\rm B\kern-.05em{\sc i\kern-.025em b}\kern-.08em T\kern-.1667em\lower.7ex\hbox{E}\kern-.125emX}}

\makeatletter
\let\old@ps@headings\ps@headings
\let\old@ps@IEEEtitlepagestyle\ps@IEEEtitlepagestyle
\def\confheader#1{%

\def\ps@headings{%
\old@ps@headings%
\def\@oddhead{\strut\hfill#1\hfill\strut}%
\def\@evenhead{\strut\hfill#1\hfill\strut}%
}%

\def\ps@IEEEtitlepagestyle{%
\old@ps@IEEEtitlepagestyle%
\def\@oddhead{\strut\hfill#1\hfill\strut}%
\def\@evenhead{\strut\hfill#1\hfill\strut}%
}%
\ps@headings%
}
\makeatother
\confheader{}%

\title{Resilient Channel Charting Under Varying Radio Link Availability\\}

\def\authorrefmark#1{\ensuremath{^{\textbf{#1}}}}
\author{
    Jonas Pirkl\authorrefmark{1},
    Jonathan Ott\authorrefmark{1},
    Maximilian Stahlke\authorrefmark{1},
    George Yammine\authorrefmark{1},
    Tobias Feigl\authorrefmark{1,}\authorrefmark{2},
    and Christopher Mutschler\authorrefmark{1}\\
    {\tt\footnotesize\{jonas.pirkl, jonathan.ott, maximilian.stahlke, tobias.feigl, christopher.mutschler\}@iis.fraunhofer.de}\\ 
    \vspace{+1mm}\\ 
    \IEEEauthorblockA{\authorrefmark{1}
        Fraunhofer Institute for Integrated Circuits (IIS), 
        Division Positioning and Networks, 
        90411 Nürnberg, Germany
        }
    \IEEEauthorblockA{\authorrefmark{2}
        Computer Science Department,
        Friedrich-Alexander-Universität Erlangen-Nürnberg, 
        91052 Erlangen, Germany 
    }
}

\begin{acronym}[tdma]
    \acro{csi}[CSI]{channel state information}
    \acro{nn}[NN]{neural network}
    \acro{los}[LOS]{line-of-sight}
    \acro{nlos}[NLOS]{non-line-of-sight}
    \acro{pdr}[PDR]{pedestrian dead reckoning}
    \acro{cc}[CC]{channel charting}
    \acro{fp}[FP]{fingerprinting}
    \acro{mpc}[MPC]{multi path component}
    \acro{mae}[MAE]{mean average error}
    \acro{cir}[CIR]{channel impulse response}
    \acro{mse}[MSE]{mean squared error}
    \acro{adp}[ADP]{angle-delay profile}
    \acro{tdoa}[TDoA]{time-difference of arrival}
    \acro{pca}[PCA]{principal component analysis}
    \acro{rssi}[RSSI]{Received Signal Strength Indicator}
    \acro{mlp}[MLP]{multi-layer perceptron}
    \acro{ml}[ML]{machine learning}
    \acro{adapos}[AdaPos]{Adaptive Positioning}
    \acro{sota}[SOTA]{state-of-the-art}
    \acro{cnn}[CNN]{Convolutional Neural Network}
\end{acronym}

\begin{document}
\maketitle

\begin{abstract}
\Ac{cc} has become a key technology for RF-based localization, enabling unsupervised radio fingerprinting, even in non line of sight scenarios, with a minimum of reference position labels. 
However, most \ac{cc} models assume fixed-size inputs, such as a constant number of antennas or channel measurements. 
In practical systems, antennas may fail, signals may be blocked, or antenna sets may change during handovers, making fixed-input architectures fragile. 
Existing radio-fingerprinting approaches address this by training separate models for each antenna configuration, but the resulting training effort scales prohibitively with the array size.

We propose \ac{adapos}, a \ac{cc} architecture that natively handles variable numbers of channel measurements. 
\ac{adapos} combines convolutional feature extraction with a transformer-based encoder using learnable antenna identifiers and self-attention to fuse arbitrary subsets of CSI inputs.
Experiments on two public real-world datasets (SISO and MIMO) show that \ac{adapos} maintains state-of-the-art accuracy under missing-antenna conditions and replaces roughly 57 configuration-specific models with a single unified model.
With \ac{adapos} and our novel training strategies, we provide resilience to both individual antenna failures and full-array outages.
\end{abstract}

\section{Introduction}

\begin{figure}
    \centering
    \includegraphics[trim={19 14 19 15},clip,width=1\linewidth]{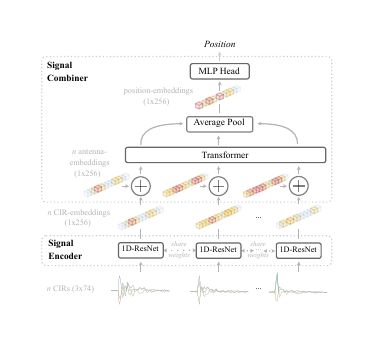}
    \vspace{-0.6cm}
    \caption{Overview of the \ac{adapos} architecture. 
        Information flows from bottom to top. 
        We first pass individual channel impulse responses (CIRs) through a signal encoder (1D-ResNet) that creates one embedding per \ac{cir}. 
        Next, the signal combiner adds an antenna embedding to each \ac{cir} embedding so the model identifies the source antenna. 
        A transformer encoder then combines the set of \ac{cir} embeddings through self-attention. 
        Its output is average-pooled and passed through an MLP head to return the CC pseudo 2D positions.}
    \label{fig:overview}
\end{figure}

RF-based positioning becomes increasingly important for applications such as smart factories, traffic hotspot identification, and emergency services~\cite{Laoudias2018}. 
Traditional localization methods, such as \ac{tdoa}, require \ac{los} signals and degrade in the presence of multipath propagation~\cite{Dardari2015}. 
In indoor environments, where non-line-of-sight (NLOS) conditions and signal blockages occur frequently, these methods often fail to provide reliable accuracy~\cite{Dardari2015}. 
Radio fingerprinting offers an alternative by mapping measured radio features to positions and has shown promising performance in indoor scenarios~\cite{Dardari2015}. 
However, fingerprinting depends on large labeled datasets, and collecting such data remains time-consuming and expensive~\cite{Dardari2015}. 

\Ac{cc} addresses this limitation by learning a low-dimensional representation of \ac{csi} that reflects the spatial neighborhood structure without requiring reference location labels~\cite{cc, pdr_cc, triplet_cc, adp-metric}. 
\Ac{cc} maps the \ac{csi} manifold to a latent space using pseudo-distance metrics, enabling scalable, unsupervised RF localization~\cite{cc}. 
Most fingerprinting~\cite{fontaine_ultra_2023} and \ac{cc} approaches~\cite{triplet_cc, poeggel2025passivechannelchartinglocating} assume a fixed input size, meaning that the number of antennas or channel measurements remains constant. 
In practice, however, antenna availability changes: antennas may be powered down to save energy~\cite{powersaving}, signals may be blocked~\cite{blockage}, or hardware failures may occur~\cite{outage}. 
These situations lead to variable radio link availability, which breaks fixed-input models. 
Existing solutions address this by training separate models for different antenna configurations and selecting the appropriate model during inference~\cite{fontaine_ultra_2023}. 
Since the number of antenna subsets grows exponentially with the array size ($\mathcal{O}(2^n)$), training and maintaining configuration-specific models quickly becomes infeasible.

Our paper addresses these limitations. 
We propose \ac{adapos} (see Fig.~\ref{fig:overview}), a \ac{cc} architecture that handles a variable numbers of channel measurements natively. 
\ac{adapos} uses a shared \ac{cnn} to extract \ac{cir} embeddings, augments them with learnable antenna identifiers, and employs a transformer-based encoder to fuse arbitrary subsets of \acp{cir} through self-attention~\cite{trafo}. 
This allows a single model to adapt to changing antenna configurations without retraining. 
We further introduce training strategies that simulate antenna dropouts and improve robustness when only few signals are available. 
These strategies also allow existing models to handle missing antennas without architecture changes. 
In this way, \ac{adapos} avoids the exponential growth of configuration-specific models and provides a scalable solution for large antenna arrays.
We evaluate \ac{adapos} on two real-world datasets ($B_1 = 50\,\text{MHz},\, f_{c1} = 1.3\,\text{GHz}$ and 
 $B_2 = 100\,\text{MHz},\, f_{c2} = 3.7\,\text{GHz}$) and test a range of missing-antenna scenarios. The experiments show that \ac{adapos} maintains state-of-the-art performance (maximum of 6~\si{cm} deviation) across varying antenna configurations and replaces roughly 57 configuration-specific models with one unified model while keeping nearly unchanged error performance.

The remainder of this paper is organized as follows. 
Sec.~\ref{sec:relwork} discusses related work. 
Sec.~\ref{sec:method} describes the \ac{adapos} architecture and training strategies.
Sec.~\ref{sec:experimental_setup} presents the datasets and evaluation setup, and
Sec.~\ref{sec:eval} reports the experimental results, before Sec.~\ref{sec:conclusion} concludes.

\section{Related Work}\label{sec:relwork}

\ac{cc} was introduced as a scalable alternative that removes the dependence on labeled reference positions. 
\ac{cc} learns a lower-dimensional representation of \ac{csi} that preserves the spatial neighborhood structure of the high-dimensional space~\cite{cc}.

Early work uses classical dimensionality-reduction methods such as PCA \cite{cc}, Sammons Mapping \cite{cc}, autoencoders \cite{cc} or t-SNE \cite{mp-cc}.
More recently, neural-network-based methods use pseudo-distance metrics, that \ac{cc} uses to map high-dimensional \ac{csi} to a lower-dimensional latent space, enabling unsupervised localization.
These pseudo-distances can be derived directly from \ac{csi}~\cite{cosine_sim, geodesic} or from auxiliary modalities such as timestamps~\cite{triplet_cc} or pedestrian dead reckoning (PDR)~\cite{pdr_cc}. 
Recent metrics explore using geodesic distances \cite{geodesic} or fusing multiple metrics~\cite{adp-metric} to improve quality of the resulting channel chart.

Popular architectures are based on neural networks and include, for instance, Siamese networks~\cite{siamese-cc, stahlke2024velocity, adp-metric, poeggel2025passivechannelchartinglocating} and triplet networks~\cite{triplet_cc}. 
Triplet networks \cite{triplet} optimize a contrastive objective, where samples close in the data space are pulled together in the latent space, while distant samples are pushed apart.
In \ac{cc}, temporal proximity can serve as a proxy for spatial proximity: samples recorded within short intervals are assumed to be close in distance, while samples separated by longer intervals are expected to be farther apart because the person has moved more~\cite{triplet_cc}.
Siamese networks~\cite{koch2015siamese} are trained to predict positions by minimizing the mismatch between the Euclidean distance of two predicted positions and the pseudo-distance between the corresponding samples~\cite{siamese-cc}.
These parametric models are efficient at inference and can learn complex \ac{csi}–distance relationships. 

However, \ac{nn}-based \ac{cc} methods typically assume the same antennas are available at training and inference~\cite{poeggel2025passivechannelchartinglocating, geodesic, adp-metric}.
When antenna availability changes, these models degrade or fail. 
A common workaround is to train a separate model for each antenna subset and select the appropriate model during inference~\cite{fontaine_ultra_2023}.
While feasible in small systems, this approach does not scale. 
The number of possible antenna subsets grows exponentially ($\mathcal{O}(2^n)$), making it impractical for arrays with many elements.

In contrast to existing approaches, we provide a single model, that handles any antenna configuration natively. 
By designing a model that processes variable numbers of \acp{cir} and remains robust to missing antennas, we remove the need for configuration-specific models and address the scalability challenges inherent in existing \ac{cc} and fingerprinting solutions.

\section{Method}\label{sec:method}
We first introduce the \ac{adapos} architecture and then explain how we train and handle varying antenna availability in \ac{cc}.

\subsection{Network Architecture}\label{subsec:loss}
To accommodate variable numbers of received signals, i.e. radio links, we propose our flexible transformer-based model called \ac{adapos}. 
The architecture consists of two main components: a signal encoder and a signal combiner, see
Fig.~\ref{fig:overview}.

The \textbf{Signal Encoder} (lower part in Fig.~\ref{fig:overview}) processes individual \acp{cir} and creates a compact embedding that captures their most relevant features. 
It is implemented as a lightweight 1D-ResNet whose weights are shared across all \ac{cir} inputs, which reduces the number of parameters and improves generalization. 
The resulting embeddings are passed directly to the signal combiner. 

The \textbf{Signal Combiner} (upper part in Fig.~\ref{fig:overview}) fuses an arbitrary set of embeddings and predicts the final position. 
It supports any number of inputs and learns how much each contributes to the final estimate. 
Its core mechanism is a self-attention layer~\cite{trafo}, which dynamically extracts relevant information from all available embeddings. 
For each input embedding, the attention module forms a weighted sum of all embeddings, where the weights are derived from learned similarity scores. 
As the weighted sum has a fixed output size independent of the number of inputs, the model naturally handles varying antenna counts. 
However, this operation is permutation-invariant. 
To retain information about the origin of each \ac{cir}, we add a learnable antenna-specific embedding to each \ac{cir} embedding, allowing the model to identify the source antenna. 
This approach improves performance under dynamic antenna availability as the attention weights depend directly on the available inputs, in contrast to \acp{cnn} or \acp{mlp}, which rely on fixed receptive fields or fixed input dimensions. 
We implement the signal combiner as a compact transformer encoder~\cite{trafo} with a hidden dimensionality of 256, 8 attention heads, and a feed-forward layer of size 1024, using a total of 3 encoder layers. 
The transformer output is average-pooled and passed through an \ac{mlp} head to predict the 2D position.

\subsection{Model Training}\label{subsec:training}
We adopt a Siamese training strategy to project high-dimensional \ac{csi} into a 2D coordinate space. 
2 samples pass through the same network, and the loss enforces that the Euclidean distance between the predicted positions matches the pseudo-distance derived from the \ac{cc} metric.
The loss function is defined as
\begin{equation}
\label{eq:cc_dist}
\mathcal{L}_\theta(x_n, x_k, d_{n,k}) = (d_{n,k}-|f_{\theta}(\ve{x}_n)-f_{\theta}(\ve{x}_k)|_2)^2,
\end{equation}
where $f_{\theta}$ denotes the neural network with parameters $\theta$, $x_n$ and $x_k$ are two \ac{cir} samples, and $d_{n,k}$ their pseudo-distance.

\section{Experimental Setup}\label{sec:experimental_setup}
We first introduce the training strategies employed for \ac{adapos} and the ResNet baseline.
Next, we summarize the datasets and preprocessing steps, followed by the training configuration used in all experiments.

\subsection{Training Strategies}\label{sec:training_strategies}
We investigate two training strategies for both \ac{adapos} and a ResNet baseline architecture: (1) Fixed-N and (2) Random-N. 
Fixed-N trains the models with a fixed number of antennas $n_t$, ranging from 1 to $a_\text{max}$, where $a_\text{max}$ is the maximum number of antennas available in the dataset. 
In each batch, we select $n_t$ random antennas and pass them through the model. 
Random-N first samples the number of antennas $n_t$ uniformly for every batch ($n_t \sim \mathcal{U}(1, a_{\max})$), and then select $n_t$ random antennas as inputs. 
In general, for optimization, we use an AdamW optimizer~\cite{admaw} and apply a linear warmup schedule~\cite{trafo} for the learning rate to stabilize training during the initial epochs, and a weighted \ac{mse} loss, Sec. \ref{subsec:loss}.

\subsection{Dataset}\label{sec:dataset}
To evaluate the proposed approach, we use 2 real-world datasets: the Dichasus dataset~\cite{dichasus} and the Fraunhofer 5G dataset~\cite{5g-dataset}. 
These datasets represent diverse environments and antenna configurations, providing a robust basis for evaluating model robustness under varying antenna availability.

\textbf{Dichasus Dataset}~\cite{dichasus} covers a MIMO setup recorded in an industrial environment using four antenna arrays, each consisting of eight antenna elements. 
The system uses a 50.056\,MHz bandwidth at a center frequency of 1.272\,GHz.
The dataset provides \ac{csi} sampled at 20.83\,Hz, which we transform into the time domain and normalize to the range $[0,1]$. 
Each \ac{cir} has a resolution of 80 samples and 3 channels: real part, imaginary part, and the absolute value, similar to~\cite{cir_abs}.
We employ the current \ac{sota} for antenna arrays as a distance metric, which is a geodesic fusion of a time-based and \ac{adp}-based metric~\cite{adp-metric}. We train on 60,507 samples of dichasus-cf02 and dichasus-cf04, and test on 23,478 samples of dichasus-cf03.

\textbf{Fraunhofer 5G Dataset} covers a SISO setup recorded in a simulated industrial warehouse environment using a 5G downlink system with six receiving antennas. 
The radio operates at 3.7\,GHz with a bandwidth of 100\,MHz and a transmit power of 20\,dBm.
Data is recorded over an area of approximately $400\,\mathrm{m}^2$ at 6.6\,Hz. 
The training set contains 18,706 samples and the test set contains 15,722 samples.
This time we employ the CIRA metric~\cite{geodesic}, which also works with single antennas.

\subsection{Training Configuration}\label{sec:training_config}
We train and evaluate the models under various antenna configurations to assess robustness to missing antennas.
Specifically, we train models for every training strategy as described in \Cref{sec:training_strategies}. We then evaluate each of the models for every possible count of antennas for the corresponding dataset.
We include a ResNet-based architecture~\cite{poeggel2025passivechannelchartinglocating} as a baseline to compare how the training strategies affect transformer- and \ac{cnn}-based models.
For the ResNet baseline, missing antennas are handled by zeroing out the corresponding input channels. 
In contrast, \ac{adapos} is designed to process variable-sized inputs directly. 
The learnable antenna embeddings allow the model to identify the source antenna for each \ac{cir}, so missing antennas are simply omitted from the input set.

\begin{table}[!bp]
    \centering
    \caption{Baseline results, averaged across all antenna combinations with the same number of visible antennas.}
    \begin{tabular}{@{}ccc@{}}
    \multicolumn{1}{l}{No.} & \multicolumn{1}{l}{No.} & \multicolumn{1}{l}{MAE} \\ 
    \multicolumn{1}{l}{antennas} & \multicolumn{1}{l}{models} & \multicolumn{1}{l}{{[}m{]}} \\
    \midrule
    2                  & 15                                   & 1.43        \\
    3                  & 20                                   & 1.39        \\
    4                  & 15                                   & 1.39        \\
    5                  & 6                                    & 1.39        \\
    6                  & 1                                    & 1.37        \\ 
    \end{tabular}
    \label{tab:fr1-baselines}
\end{table}

\begin{figure}[!tp]
   \centering
    \includegraphics[trim={0 0 0 0.55cm}, clip, width=1\linewidth]{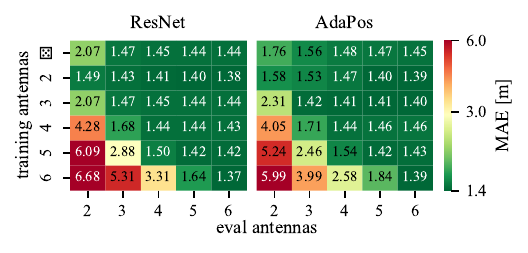}
    \vspace{-1cm}
    \caption{MAE of ResNet (left), AdaPos (right) on the Fraunhofer 5G dataset.}
    \label{fig:fr1-cc-resnet}
\end{figure}

\begin{figure*}[!tp]
   \centering
    \includegraphics[trim={0 0 0 0.55cm}, clip, width=1\linewidth]{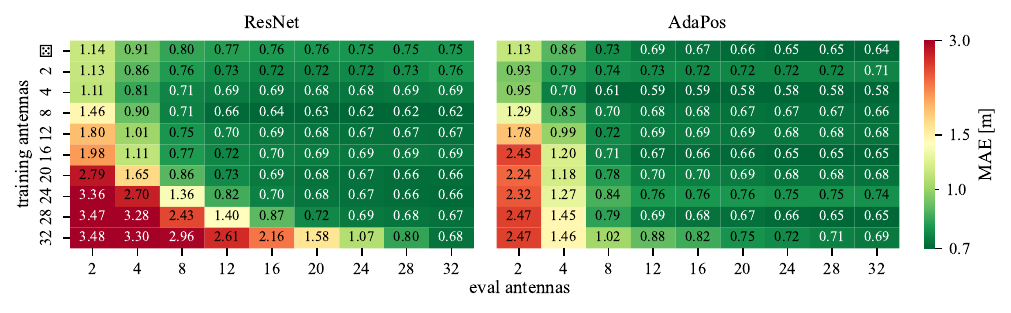}
    \vspace{-1cm}
    \caption{MAE of ResNet (left), AdaPos (right) on the Dichasus dataset, w. single antennas masked.}
    \label{fig:dichasus_cc}
\end{figure*}

\begin{figure*}[!tp]
   \centering
    \includegraphics[trim={0 0 0 0.55cm}, clip, width=1\linewidth]{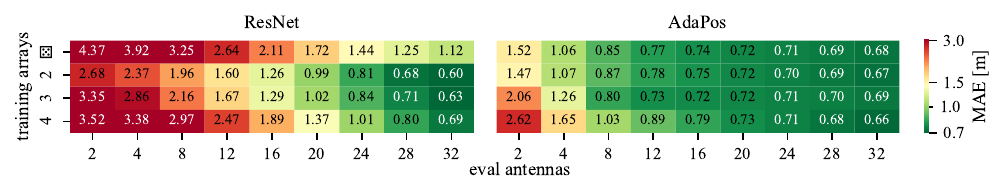}
    \vspace{-1cm}
    \caption{ResNet (left), AdaPos (right) trained with full arrays masked and evaluated with single elements missing.}
    \label{fig:dichasus_taes_cc_both_models}
    \vspace{-0.4cm}
\end{figure*}

\begin{figure}[!tp]
   \centering
    \includegraphics[trim={0 0 0 0.55cm}, clip, width=\linewidth]{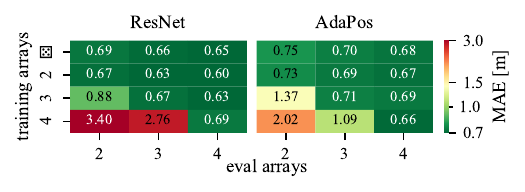}
    \vspace{-1cm}
    \caption{MAE of ResNet (left), AdaPos (right) on the Dichasus dataset, w. arrays masked.}
    \label{fig:dichasus_array}
\end{figure}

\begin{figure}[!tp]
   \centering
    \includegraphics[trim={0 0 0 0.55cm}, clip, width=1\linewidth]{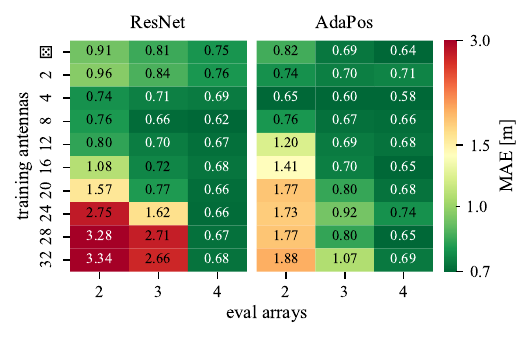}
    \vspace{-1cm}
    \caption{ResNet (left), AdaPos (right) trained with single elements masked and evaluated with full arrays missing.}
    \label{fig:dichasus_tsea_cc_both_models}
    \vspace{-0.4cm}
\end{figure}

\section{Evaluation}\label{sec:eval}

Benchmarking all antenna combinations is computationally infeasible. 
Thus, for each evaluation antenna count $n_e$, we randomly select the missing antennas in every batch to obtain a representative estimate of the overall performance. 
For a fair comparison, we use the same random antenna selections for all methods. 
We omit the case $n_e = 1$, as neither training nor evaluating single-antenna signals yields meaningful results due to the lack of spatial diversity, and \ac{adapos} relies on attention weights that compare multiple embeddings, that are undefined for a single input.
To compute the localization error, we apply the optimal affine transformation to map the channel chart from pseudo-coordinates to real-world coordinates.

\subsection{Evaluation on the Fraunhofer 5G Dataset}\label{sec:eval:5g}
As a baseline, we train 57 configuration-specific models, one for every antenna subset containing 2 to 6 antennas. 

Table~\ref{tab:fr1-baselines} shows that performance does not vary significantly across different antenna combinations.
This indicates that the accuracy is mainly limited by the \ac{cc} metric rather than by the number of available antennas.
Fig.~\ref{fig:fr1-cc-resnet} compares the ResNet and \ac{adapos} models. 
The plots show the MAE for different pairs of ($n_t$, $n_e$), where $n_t$ is the number of antennas during training and $n_e$ is the number available during inference. 
We also evaluate the Random-N strategy ($n_t \sim \mathcal{U}(1, a_{\max})$), see the dice symbol in Fig.~\ref{fig:fr1-cc-resnet}.
As expected, reducing the number of inference antennas ($n_e$) generally decreases positioning accuracy. 
However, increasing the variability during training, i.e., reducing $n_t$, substantially improves robustness. 
For example, with only 2 antennas available at inference time, the MAE improves by up to 78\% (down to 1.49\,m) when training with $n_t = 2$.
In certain settings, the ResNet baseline yields a marginally lower MAE.
We believe this is due to the relatively small number of antenna combinations, e.g., only 15 for $n_t = 2$, that increases the likelihood that ResNet encounters all combinations during training at every position.
\ac{adapos} achieves markedly more stable performance across mismatched ($n_t$, $n_e$) pairs.
In particular, \ac{adapos} degrades insignificant when the number of antennas during inference is lower than during training ($n_e < n_t$).

Overall, a single \ac{adapos} model effectively replaces all 57 configuration-specific baselines. 
Even in the most challenging case ($n_t = 2$, $n_e = 2$), MAE differs only by 6\,cm. 
In the best case ($n_t = 2$, $n_e = 6$), MAE differs by only 1\,cm.

\subsection{Evaluation on the Dichasus Dataset}
We evaluate two scenarios: (1) missing individual antenna elements within each array and (2) missing entire antenna arrays. 
As the dataset consists of 4 arrays with 8 elements each, testing all antenna combinations is infeasible. 
So, we treat $n_e = 32$ (or $n_e = 4$ when evaluating missing arrays) as a practical lower bound.

\textbf{Single Antenna Element.} We consider 2 categories of models. The first category is trained with single-element masking ($a_{\max} = 32$), see Fig.~\ref{fig:dichasus_cc}. 
The second category is trained with full-array masking ($a_{\max} = 4$), see Fig.~\ref{fig:dichasus_taes_cc_both_models}.
Models trained with single-element masking show trends similar to the Fraunhofer dataset: \ac{adapos} outperforms ResNet whenever $n_e < n_t$. 
Also, the ResNet model struggles when only a few antennas are available ($n_e \leq 12$), whereas \ac{adapos} remains robust across all training strategies.
Interestingly, training on 4 single-element masks ($n_t = 4$) yields the best-performing \ac{adapos} model, reflecting that optimal training strategies are dataset-dependent.
When training with entire arrays masked, the ResNet model performs poorly overall. 
\ac{adapos} improves over ResNet but still falls short of the model
trained with single-element masking. 
So, using \ac{adapos} with single-element masking improves generalization, even when entire arrays fail.

\textbf{Antenna Arrays.} We evaluate the performance when whole arrays are missing, see Fig.~\ref{fig:dichasus_array}. 
Here, the ResNet baseline performs slightly better than \ac{adapos} in most settings except when all arrays are present ($n_e = 4$). 
However, when training with single-element masking rather than array masking, see Fig.~\ref{fig:dichasus_tsea_cc_both_models}, the best \ac{adapos} model ($n_t = 4$) outperforms the best array-masked baseline (ResNet with $n_t = 2$). 
So, training with single-element masking yields a model that is robust not only to isolated antenna failures but also to complete array outages.

\section{Conclusion} \label{sec:conclusion}
We presented \ac{adapos}, a transformer-based \ac{cc} architecture that natively handles variable radio link availability. 
Our approach eliminates the need for exponentially many configuration-specific models, an otherwise intractable requirement for large arrays, while maintaining \ac{sota} localization accuracy.
Experiments on 2 real-world datasets show that a single \ac{adapos} model replaces a large number of configuration-specific baselines.
On SISO 5G \ac{csi}, we substituted 57 dedicated models with only a marginal accuracy loss (at most 6\,cm MAE) and consistently generalizes better across mismatched training-inference antenna configurations.
On MIMO \ac{csi}, \ac{adapos} trained with single-element masking provides strong resilience against both individual element failures and full-array outages, outperforming models trained solely on array masking.
The optimal masking strategy is dataset-dependent, but the Random-N strategy yields robust performance without per-dataset tuning.

\section{Acknowledgments}
This work was supported by the Federal Ministry of Education and Research of Germany in the programme of “Souverän. Digital. Vernetzt.” joint project 6G-RIC (16KISK020K).

\AtNextBibliography{\small}
\printbibliography

\end{document}